# Reimagining Assessment in the Age of Generative AI: Lessons from Open-Book Exams with ChatGPT

*Qusay H. Mahmoud*
Faculty of Engineering and Applied Science
Ontario Tech University, Oshawa, Ontario L1G 0C5 Canada
qusay.mahmoud@ontariotechu.ca

*This paper was accepted to the 2026 Canadian Engineering Education Association Conference, but was later withdrawn prior to publication.*

**Abstract** – *Generative AI systems such as ChatGPT challenge traditional assumptions about academic assessment by enabling students to generate explanations, code, and solutions in real time. Rather than attempting to restrict AI use, this study investigates how students actually interact with such systems during formal evaluation. Engineering students were permitted to use ChatGPT during take-home open-book exams and were required to submit interaction transcripts alongside exam solutions. This provided direct observational evidence of reasoning processes rather than relying on self-reported behavior. Qualitative analysis revealed three progressive patterns of use: answer retrieval, guided collaboration, and critical verification. While some students initially copied questions verbatim and received generic responses, many refined prompts iteratively and tested outputs. Some of the strongest evidence of reasoning appeared when students evaluated incorrect or incomplete AI responses, revealing evaluative reasoning through debugging, comparison, and justification. The presence of generative AI shifted the cognitive task of assessment from producing solutions to assessing solution validity. The findings suggest that, in AI-mediated assessment environments, correctness of final answers alone may no longer provide sufficient evidence of comprehension. Instead, competencies such as prompt formulation, verification, and judgment become visible indicators of learning. Transparent integration of AI appeared to reduce focus on rule avoidance and promote self-regulation. Assessments should evolve to evaluate reasoning about solutions rather than independent solution production. Generative AI therefore does not invalidate assessment but has the potential to expose deeper forms of understanding aligned with professional practice.*

## 1. Introduction

The rapid adoption of generative artificial intelligence (AI) systems such as ChatGPT is reshaping higher education at a pace rarely seen with previous educational technologies [1,8]. Students can now generate explanations, debug programs, summarize concepts, and explore alternative solutions in real time through conversational interaction. This development challenges long-standing assumptions about assessment: if answers can be produced instantly, what exactly are examinations measuring?

Historically, open-book assessments were designed to shift evaluation away from memorization toward reasoning and application [9]. However, large language models introduce a fundamentally different dynamic. Instead of merely retrieving information, students can collaborate with an AI system that synthesizes responses, generates code, and adapts explanations to student queries. As a result, the boundary between consulting resources and outsourcing cognition becomes increasingly unclear. Educational institutions have responded inconsistently, some restrict AI use, while others attempt to detect it [4, 5, 6], but both approaches struggle to address a central pedagogical question: how do students actually use AI when it is openly available during assessment? This question is also connected to recent research on generative AI, assessment, self-regulated learning, and evaluative judgment in higher education [11–14]. In AI-mediated learning environments, students are not only retrieving information but also planning prompts, monitoring responses, identifying uncertainty, and deciding whether outputs are trustworthy. These behaviors align with emerging work suggesting that generative AI can support learning when students actively regulate, evaluate, and reflect on AI-generated outputs rather than accepting them passively [11,13,14]. From this perspective, effective AI use depends not only on access to the tool but also on students' ability to evaluate the quality of AI-generated responses and justify evaluation decisions, a capacity closely related to evaluative judgment in assessment [12].

Unlike traditional resources consulted in open-book exams, generative AI produces synthesized responses rather than directing students to information sources. As a result, the task shifts from locating knowledge to evaluating generated explanations, making open-book assessment an appropriate setting for observing how reasoning changes in AI-mediated environments.

Rather than treating generative AI as misconduct or attempting to eliminate it from assessment environments, this study adopts a different stance: AI use is acknowledged as inevitable and therefore should be studied in authentic academic settings. Engineering students were permitted to use ChatGPT during take-home open-book exams and were required to submit interaction transcripts alongside exam solutions. This transformed the assessment itself. Instead of evaluating only final answers, the study gained visibility into students' reasoning processes, including problem framing, output interpretation, result validation, and negotiation of trust with the AI system.

The resulting dataset provides empirical insight into how students actually use generative AI during assessment, revealing distinct interaction patterns and verification behaviors. In many cases, learning occurred not from the AI's correctness but from its limitations: students identified errors, challenged responses, and developed verification strategies.

This paper examines how engineering students employed ChatGPT during open-book examinations and what this reveals about assessment design in the age of generative AI. Specifically, three questions are investigated:

*RQ1) How do students interact with generative AI during formal assessments?*

*RQ2) What learning behaviors, challenges, and ethical considerations emerge from this interaction?*

*RQ3) How should assessments evolve when intelligent tools can participate in the problem-solving process?*

By grounding the discussion in real interaction data rather than speculation, this work contributes evidence-based guidance for educators navigating AI-integrated learning environments. The findings suggest that generative AI does not simply threaten academic integrity [4]; instead, it exposes new competencies—prompt formulation, critical verification, and epistemic judgment—that traditional exams rarely measure but modern education increasingly requires.

This paper makes the following contributions:

1. A validation interaction progression model.
2. A chat-transcript-based assessment methodology.
3. Evidence that verification behaviors are associated with deeper reasoning.
4. A design framework for AI-integrated engineering assessment.

The rest of this paper is organized as follows. Section 2 reviews related work on generative AI and assessment in higher education. Section 3 describes the educational context and assessment design. Section 4 outlines the research methodology. Section 5 presents the findings from analysis of student–AI interaction transcripts. Section 6 discusses implications for learning and assessment. Section 7 provides design recommendations for AI-integrated evaluation, and Section 8 concludes the paper with ideas for future work.

## 2. Background and Related Work

The emergence of generative AI in education builds upon a long history of intelligent tutoring systems and digital learning aids, but large language models represent a qualitative shift rather than an incremental improvement [1,3]. Earlier tools automated feedback within narrowly defined domains, whereas systems such as ChatGPT provide conversational explanations across diverse topics, generate code, and adapt responses dynamically. This flexibility has led students to adopt AI not merely as a reference source but as an interactive cognitive partner capable of participating in problem solving [1,8].

Recent discussions in higher education have focused on whether generative AI should be restricted, detected, or incorporated into academic work. Concerns commonly center on academic integrity, originality, and the possibility that students may bypass meaningful learning by delegating tasks to AI. Because AI-generated text is original rather than copied, traditional plagiarism detection methods are ineffective [4,5], and attempts at automated detection remain unreliable. Consequently, many institutions face a practical dilemma: enforcement is difficult, yet ignoring the technology risks undermining assessment validity [6].

At the same time, emerging research suggests generative AI can support learning when students engage critically with AI-generated outputs rather than accepting them passively [11-14]. Students frequently use AI tools to obtain alternative explanations, clarify misunderstandings, explore examples, and receive immediate feedback beyond lecture materials. In this role, AI resembles an always-available tutor [1], capable of adapting explanations to student needs. However, the educational value of these systems depends strongly on how students interact with these systems. Recent studies emphasize that meaningful learning requires active prompting, verification, reflection, and evaluative judgment rather than passive acceptance of generated responses [11-14]. This has elevated prompt engineering from a technical curiosity to a form of academic literacy [7], requiring students to communicate intent, constraints, and expectations effectively.

Assessment design is therefore becoming the central issue. Traditional exams often assume the student works independently [9,10] to produce an answer. When AI can generate answers instantly, such assessments risk measuring tool usage rather than understanding. In response, educators have begun exploring alternative strategies: emphasizing reasoning, requiring reflection [9], incorporating oral explanation, or documenting the problem-solving process. Instead of asking whether AI should be allowed, the more productive question becomes what skills should be evaluated in an AI-augmented environment.

Despite growing debate, empirical evidence remains limited regarding how students actually use generative AI during formal assessments. Much of the existing literature focuses on institutional policy, academic integrity concerns, student perceptions, or theoretical implications of AI in education [4-8,11-14]. Fewer studies directly observe authentic student–AI interaction during real assessment tasks, particularly in ways that make reasoning processes visible.

This study contributes to that gap by analyzing complete ChatGPT interaction transcripts collected during open-book examinations. By examining prompts, revisions, validation attempts, and student reflections, the analysis moves beyond theoretical concerns to observable practices. The goal is not simply to evaluate the effectiveness of AI assistance, but to understand how the presence of an intelligent collaborator reshapes the nature of assessment and the competencies it reveals.

## 3. Context and Assessment Design

This study was conducted across one undergraduate and two graduate-level engineering courses: Introduction to Programming for Engineering (Summer 2023), Network Computing (Winter 2023), and Programming Methodology and Abstraction for Engineers across five offerings (Fall 2023, Winter 2024, Fall 2024, Winter 2025, and Fall 2025). Across these courses, students developed programming, problem-solving, and network computing skills. Students came from diverse engineering backgrounds, many without formal computer science training, and the courses emphasized practical reasoning, debugging, and conceptual understanding rather than memorization.

### 3.1 Rationale for Allowing AI During Exams

With the widespread availability of generative AI tools, students were already informally using conversational systems to support learning outside the classroom. Rather than attempting to prohibit this behavior, the assessment design intentionally incorporated AI as an authorized resource. The goal was not to simplify the exam but to observe and evaluate how students reason when an intelligent assistant is available.

Traditional open-book exams assume consultation of static resources; generative AI instead produces synthesized responses, shifting emphasis toward reasoning processes.

Students were permitted to use ChatGPT during take-home exams under two explicit conditions:

1. Students were required to validate and understand any AI-generated content used in their submissions.
2. Students had to submit the full ChatGPT interaction transcript alongside exam solutions.

This requirement transformed AI use from hidden assistance into observable academic behavior. The transcript acted as a record of thinking—revealing how students framed problems, revised questions, evaluated outputs, and responded to uncertainty.

### 3.2 Dual-Artifact Assessment Model

The assessment evaluated two complementary artifacts:

1. Final Solutions: the completed answers submitted by students.
2. Interaction Logs: the ChatGPT conversation transcripts documenting the problem-solving process.

Rather than penalizing AI usage, the design encouraged responsible engagement. Students were informed that the transcript would not be graded for correctness of AI output but for evidence of reasoning, validation, and thoughtful interaction. This approach positioned AI as a cognitive tool similar to a calculator or development environment, useful but requiring interpretation.

### 3.3 Intended Learning Behaviors

The design aimed to promote three learning behaviors:

1. Prompt formulation: articulating questions precisely enough to obtain meaningful responses.
2. Critical verification: testing, debugging, and cross-checking AI outputs.
3. Reflective judgment: deciding when to trust, modify, or reject AI suggestions.

By embedding these behaviors into the assessment environment, the exam became an opportunity to evaluate higher-order competencies rarely visible in traditional testing. Instead of asking whether students could produce answers independently, the assessment examined student reasoning while working alongside an intelligent system. This context provided a natural setting to observe authentic AI-assisted problem solving and enabled analysis of real student decision-making rather than hypothetical or self-reported behavior.

## 4. Methodology

This study employed a qualitative empirical approach to examine how students interact with generative AI during formal assessment. By collecting complete ChatGPT interaction transcripts alongside exam submissions, the study obtained direct evidence of student reasoning processes rather than relying solely on surveys or self-reporting. The study was guided by an interpretivist qualitative research perspective focused on understanding how students made sense of and interacted with generative AI during authentic assessment tasks. The goal was not to measure causal learning outcomes, but to examine observable reasoning behaviors, validation practices, and patterns of interaction emerging through student–AI dialogue.

### 4.1 Participants and Data Sources

Data were collected from engineering students across multiple offerings of three courses: Network Computing (graduate, Winter 2023), Introduction to Programming for Engineering (undergraduate, Summer 2023), and *Programming Methodology and Abstraction* (5 offerings), 94 students participated in take-home open-book exams where ChatGPT use was explicitly permitted. The dataset contained approximately 94 interaction transcripts comprising several thousand prompts and responses.

Two primary data sources were analyzed:

1. ChatGPT interaction transcripts: full conversation logs submitted with each exam.
2. Post-exam reflections: feedback describing students' experiences, strategies, and perceptions of AI use.

The transcripts provided step-by-step records of student–AI interaction, including prompts, revisions, follow-up questions, and corrections. The reflections complemented the interaction transcripts by capturing students' intentions, challenges, and ethical perspectives. In addition to supporting thematic coding, the reflections were used to triangulate interpretation by comparing observed interaction behaviors with students' own explanations of ChatGPT use.

### 4.2 Ethical Considerations

This study adhered to ethical research principles and used pre-existing data originally collected for educational purposes. Where records contained student identifiers, such identifiers were removed prior to analysis, and the data were not linked to or correlated with student identities. The study did not involve intervention, manipulation, or identification of individual participants, and findings are reported only in aggregate form.

### 4.3 Analytical Approach

Because the goal was to understand behaviors rather than measure performance differences, reflexive thematic analysis, as described by Braun and Clarke [15], was used. The analysis proceeded in three stages.

**Stage 1: Interaction Pattern Identification**

All transcripts were reviewed to identify recurring interaction behaviors. Prompts were coded according to how students engaged with the AI, including:

- direct question copying
- contextualized prompting
- iterative refinement
- debugging queries
- verification attempts

These categories emerged inductively from the data rather than being predetermined.

**Stage 2: Prompt Effectiveness Classification**

Each interaction sequence was evaluated in relation to the exam task it addressed. Prompts were classified as:

- Effective: produced useful, accurate, or actionable responses with minimal correction
- Ineffective: produced irrelevant, incorrect, or overly generic responses requiring substantial revision

This classification focused not on whether the AI output itself was correct, but on whether the interaction reflected deliberate student guidance, refinement, and evaluation of the tool's responses.

**Stage 3: Thematic Interpretation**

Student reflections and interaction evidence were analyzed together to identify broader learning behaviors and challenges. Themes included:

- development of prompt strategies
- validation and debugging practices
- trust and skepticism toward AI output
- ethical concerns regarding reliance and originality

Triangulating transcripts with reflections allowed observed behavior to be interpreted in light of student intent.

### 4.4 Limitations

Students were aware that their ChatGPT transcripts would be reviewed, which may have influenced behavior toward more deliberate interaction. Additionally, the study was conducted within computing-related courses, so the types of tasks involved analytical reasoning and programming rather than purely narrative writing. The findings therefore emphasize cognitive processes related to problem solving, though many observations are transferable to other disciplines. Despite these limitations, the transcript-based methodology provides direct observational evidence of AI-assisted reasoning, offering insights not accessible through surveys or post-hoc interviews alone.

In addition, the findings are based on qualitative interpretation of interaction transcripts and therefore represent context-specific analytical interpretations rather than objective measurements of learning. The study does not claim causal relationships between AI use and learning outcomes, nor does it attempt to quantify the effectiveness of particular interaction strategies across broader populations or disciplines.

This study used participation data involving analysis of submitted coursework artifacts and anonymized reflections,

and all identifying information was removed prior to analysis.

### 4.5 Threats to Validity

Several factors may influence interpretation of the results, including:

*Behavioral reactivity*: Students were aware that their ChatGPT transcripts would be reviewed. This transparency may have encouraged more deliberate interaction, increased verification behavior, or reduced superficial reliance on AI compared to unobserved usage.

*Context specificity*: The study was conducted in computing-related engineering courses involving analytical and programming tasks. Interaction patterns may differ in disciplines emphasizing argumentative writing, memorization, or factual recall, limiting generalizability across subject areas.

Assessment format. The exams were take-home open-book assessments rather than timed in-person examinations. Students had extended time for reflection and iteration, which may promote deeper interaction behaviors than would occur under time pressure.

*AI system variability*: Generative AI systems evolve rapidly. Changes in model capability or interface behavior could influence interaction patterns, meaning results reflect observed student behavior within a particular technological context rather than characteristics of future systems.

*Interpretive analysis:* The study relies on qualitative thematic interpretation of interaction transcripts. Although coding categories were derived from recurring interaction patterns, qualitative analysis is inherently interpretive and reflects one analytical reading of the data. The findings therefore represent context-specific interpretations of student reasoning behaviors rather than objective measurements. To strengthen credibility, transcript observations were triangulated with student reflections and examined across multiple interaction examples.

## 5. Findings

Analysis of the ChatGPT transcripts showed that students used AI in distinct ways reflecting different levels of reasoning and trust. Rather than simply generating answers, interactions shifted the exam toward negotiating understanding. Three patterns emerged: answer retrieval, guided collaboration, and critical verification.

### 5.1 Answer Retrieval: AI as a Search Engine

The most immediate strategy adopted by many students was direct question copying. Students pasted exam questions verbatim into ChatGPT and used the generated response as a starting point for exam answers. This behavior resembled traditional resource consultation, similar to searching documentation or online references.

While this approach was efficient, the responses were typically generic. Explanations often captured high-level definitions but lacked the specificity required for the given problem context. Students frequently followed such responses with clarifying prompts after recognizing missing details. In several cases, some students initially assumed correctness but later revised answers after encountering contradictions in subsequent steps. These revisions were visible in the transcripts through follow-up prompts that asked ChatGPT to reconsider earlier answers, explain inconsistencies, or adapt responses to the specific constraints of the exam question.

This pattern revealed the following limitation: when students treated AI as a retrieval system, it provided broad information but did not automatically satisfy task requirements. Learning occurred primarily when students recognized the mismatch between general explanations and contextual needs.

Students also used the system to interpret assignment specifications rather than to generate solutions. Many interactions consisted of clarifying problem requirements before attempting implementation. This behavior indicates that AI support may encourage engagement with problem understanding, a stage frequently bypassed in traditional programming workflows. Table 1 summarizes the prompts usage pattern and the outcomes observed for each approach.

**Table 1**: ChatGPT prompt usage strategies observed.

| Prompt Strategy | Description | Typical Outcome Quality |
|---|---|---|
| Direct copy of question | Student pastes the exam question exactly as given. | Correct but generic answers; may miss specifics or depth required for full credit. |
| Contextualized prompt | Student adds context, constraints, or desired format (e.g., specifying language, format, or including background info). | Highly relevant and tailored responses; addresses specifics well when context is well-chosen. |
| Iterative refinement (follow-ups) | Student asks follow-up questions to clarify or improve the initial response. | Improved quality through each iteration; final answers often more accurate and clearer. |
| Paraphrasing and re-querying | Student rewords the question in a different way after an unsatisfactory answer. | Often yields additional details or corrections; helps capture nuances missed by the initial phrasing. |
| Overly broad or vague prompt | Student asks a very general question (e.g., "Explain this concept" without specifics). | Superficial and generic answers; requires further prompting to become useful. |
| Exploratory/tangential queries | Student asks related or extension questions not directly required. | Provides deeper insight or alternative perspectives; can enhance understanding but may not directly earn marks. |

### 5.2 Guided Collaboration: AI as an Interactive Partner

A second group of students engaged in iterative dialogue. Rather than accepting the first answer, students refined prompts, added constraints, and requested step-by-step explanations. These students treated the system as a collaborator that required direction. Typical interactions involved:

- requesting explanations in specific formats

- asking for examples or analogies
- narrowing scope to particular cases
- progressively clarifying misunderstandings

Through this process, responses became increasingly relevant and tailored. Students learned how to ask better questions during the exam itself, demonstrating emerging prompt engineering skills. The quality of final answers improved not because the AI was inherently better, but because students learned to guide it effectively.

In these cases, the transcript showed visible reasoning progression. Students moved from uncertainty to clarity through dialogue, suggesting that learning occurred during the interaction rather than prior to it.

Students frequently characterized this interaction as similar to consulting a tutor. As one participant wrote, ChatGPT "presents complex problems in a simplified way, along with multiple solutions and examples," while another explained that repeated questioning supported understanding of "the idea behind the code, not just how to implement it."

However, interaction was not always efficient. Some students struggled to formulate precise questions or had to restate context multiple times when responses became inconsistent. These moments often led to refinement of problem understanding, but also revealed that effective AI use required communication skill rather than simple tool access.

### 5.3 Critical Verification: AI as a Fallible Tool

The most cognitively engaged behavior appeared when students challenged AI output. Many transcripts contained explicit verification actions, including testing generated code, cross-checking explanations against course materials, and questioning inconsistent results.

Incorrect outputs played a significant role in this process. When ChatGPT produced faulty calculations or flawed logic, students often re-ran code, compared results, or reformulated questions. Rather than undermining learning, these failures prompted deeper reasoning. Students had to determine not only the correct answer but also why the AI's answer was wrong.

This behavior shifted the cognitive task from solving problems to evaluating solutions. Students effectively assumed the role of reviewer, interpreting and validating a proposed solution instead of constructing one entirely from scratch.

Students explicitly acknowledged the need for validation. Several reported that the system produced inconsistent or incorrect answers and required checking: "you need to verify each and every answer," and "it didn't provide the correct answers to all the questions… background knowledge is still necessary." These reflections show that incorrect responses often triggered deeper reasoning rather than passive acceptance.

In several transcripts, students encountered confidently incorrect outputs. For example, when asked to perform a numerical calculation, the system produced different answers when questioned repeatedly, prompting students to challenge the response and re-evaluate reasoning. These interactions required students to determine not only the correct solution but also why the AI response was unreliable, reinforcing verification as an active cognitive process rather than a final confirmation step. These transcript patterns provided the clearest evidence that students were not merely using ChatGPT to obtain answers, but were evaluating the reliability, logic, and applicability of AI-generated responses.

Some debugging interactions shifted from correcting syntax to diagnosing underlying programming concepts such as memory allocation, parameter passing, and loop logic. Students repeatedly asked the system to explain why errors occurred before accepting fixes, indicating conceptual debugging rather than procedural correction.

A recurring behavior not observed in earlier datasets was the use of ChatGPT to generate test cases prior to implementation. Rather than requesting direct solutions, students frequently asked the system to produce structured validation scenarios that could later be used to evaluate correctness. This suggests a shift from answer-seeking to evaluation planning, a core component of engineering reasoning.

### 5.4 Ethical Awareness and Self-Regulation

Student reflections revealed that the presence of AI introduced ethical considerations independent of formal rules. Some students expressed concern about relying too heavily on the system, even when permitted. Others described intentionally rewriting responses to ensure understanding rather than copying output directly.

This suggests that transparency in allowed AI use shifted focus away from rule avoidance toward self-regulation. Students were not primarily focused on whether AI use was allowed, but on whether AI use supported learning.

Students frequently framed responsible use as a learning decision rather than a rule constraint. One noted that ChatGPT "should play a complementary role… not something to rely on too much," while another described it as "a guide, not a solution."

### 5.5 Student Perceptions of AI as a Learning Partner

Students described ChatGPT as supportive but unreliable, valuing its efficiency while emphasizing the need for verification. As one noted, AI "helps point me to the right direction… but you still need to verify every answer".

These perceptions align with the behavioral patterns observed in the transcripts and are summarized in Table 2.

**Table 2**: Summary of student perceptions of ChatGPT use and their corresponding educational implications.

| Observed Theme | Typical Student Perception | Educational Interpretation |
|---|---|---|
| Efficiency | Saves time searching for information | Reduces information retrieval effort |
| Explanation | Acts like a tutor | Supports conceptual understanding |
| Incorrect outputs | Requires verification | Promotes critical reasoning |
| Over-reliance risk | Must not depend entirely on AI | Encourages self-regulation |

Access to ChatGPT did not result in uniformly higher performance. As shown in Figure 1, test scores in the Network Computing course remained comparable across pre-COVID, COVID-era, and ChatGPT-permitted semesters, suggesting that AI availability did not eliminate the need for reasoning and validation.

| Winter 2023 (ChatGPT) | Winter 2021 (virtual, covid-19) | Fall 2018 (pre-covid-19) |
|---|---|---|
| – Min: 70% | – Min: 89% | – Min: 42% |
| – Max: 97.5% | – Max: 100% | – Max: 88% |
| – Average: 83% | – Average: 93% | – Average: 76% |
| – Median: 85% | – Median: 89% | – Median: 81% |

**Figure 1**: Test scores for Network Computing course.

## 6. Discussion

Prior debates on generative AI in education have emphasized risks of misconduct or policy enforcement [4,6]. In contrast, the findings show that when AI use is transparent, assessment reveals observable reasoning behaviors rather than simple answer generation. Students who refined prompts, interrogated outputs, recalculated results, and investigated inconsistencies tended to produce more context-appropriate solutions than single-query approaches. These interactions demonstrate model-auditing behavior rather than passive verification: learners evaluated not only answers but the reasoning processes behind generated answers. Consistent with prior work [1,8], generative AI does not simply automate student work; it restructures cognitive effort during assessment. In this context, understanding is revealed through interpretation, validation, and judgment rather than independent construction alone.

### 6.1 Reasoning About Solutions as Evidence of Learning

In conventional assessments, correctness of the final answer is often treated as evidence of learning. However, the transcripts show that in AI-mediated environments, correct answers can sometimes be obtained with minimal engagement when students rely uncritically on generated output. Conversely, students who deeply engaged in questioning, debugging, and refining AI responses demonstrated strong reasoning even when initial outputs were incorrect. This separation of answer correctness from cognitive competence [9] suggests that the meaningful indicator of learning shifts from independent production to evaluative judgment.

When an intelligent system can propose solutions, the primary intellectual task becomes assessing solution validity rather than constructing them entirely from memory. Students were required to determine whether outputs were logically sound, contextually appropriate, and aligned with course concepts. This mirrors professional engineering practice [9,10], where computational tools are routinely used but must be supervised and verified. Competence, therefore, is demonstrated through interpretation, validation, and justification rather than isolated recall.

Prompt formulation emerged as a key component of this evaluative process. The progression from copied questions to structured dialogue indicates that prompting itself is a cognitive skill [7]. Students who reformulated queries with clearer constraints and contextual detail obtained more accurate and relevant responses than those issuing broad, single prompts. Crafting effective prompts required identifying knowledge gaps, specifying requirements, and anticipating ambiguity, behaviors consistent with metacognitive planning. Rather than replacing thinking, generative AI exposed weaknesses in problem understanding and rewarded clarity of reasoning.

Verification behaviors provided some of the clearest evidence of evaluative reasoning. Errors generated by ChatGPT frequently triggered deeper analysis. Students tested code, recalculated results, cross-checked explanations against course materials, and reconciled inconsistencies before accepting outputs. This aligns with theories of learning through cognitive conflict [1], where confronting plausible but incorrect information promotes conceptual refinement. In this context, the AI functioned not merely as an answer generator but as a catalyst for reasoning.

Notably, several students imposed constraints on the system at the outset of interaction, requesting hints or logical explanations rather than complete solutions. This "AI restraint" behavior reflects intentional regulation of assistance and reinforces that understanding remained central to task completion. The act of evaluating a proposed solution, deciding whether to accept, modify, or reject it, became a richer demonstration of expertise than independent production alone.

Taken together, these findings indicate that generative AI does not simply automate student work [1,8]; it restructures the locus of cognitive effort. Assessment in AI-mediated environments should therefore focus on how students reason about solutions, how they formulate questions, interrogate outputs, and justify conclusions, rather than whether they can produce answers in isolation.

### 6.2 Academic Integrity Reframed

A notable outcome was the shift in student attitudes toward integrity. Because AI use was transparent and permitted, the focus moved away from rule avoidance. Students instead evaluated whether reliance on AI diminished learning. This suggests that integrity concerns may be addressed more effectively through design [4,6] than through prohibition.

When assessments emphasize reasoning and validation, copying output offers little advantage because understanding becomes necessary to justify decisions. In such environments, academic honesty is reinforced by the structure of the task rather than enforced through surveillance.

### 6.3 Assessment Design

The findings imply that assessments should evaluate competencies revealed by AI interaction rather than attempt to exclude the technology. Specifically, the presence of AI highlights three assessable abilities:

- the ability to formulate meaningful questions
- the ability to evaluate generated solutions
- the ability to justify acceptance or rejection of results

These competencies correspond to higher-order thinking and professional practice [9]. Rather than weakening assessment, generative AI exposes aspects of understanding that traditional exams rarely capture.

### 6.4 Research Question Summary

**RQ1) Student Interaction Patterns:** Students interacted with generative AI in three progressively sophisticated ways: retrieving answers, collaborating through iterative prompting, and critically verifying outputs. Interaction quality depended less on AI capability and more on student guidance of the system.

**RQ2) Learning Behaviors and Challenges:** The presence of AI revealed observable learning behaviors including prompt refinement, debugging, and validation. Challenges arose from ambiguity and incorrect outputs, which prompted deeper reasoning rather than preventing learning. Ethical concerns shifted from rule compliance to self-regulation and responsible reliance.

**RQ3) Implications for Assessment Design:** Assessment should evaluate student reasoning about solutions rather than independent solution production. Tasks that require justification, verification, and interpretation more accurately capture understanding in AI-mediated environments.

## 7. Implications and Recommendations

The issue is not AI use, but whether assessment captures learning in its presence. Instructors can redesign tasks to leverage the behaviors it reveals. The following principles follow from observed interactions.

### 7.1 Make Reasoning Visible

When only final answers are evaluated, AI assistance obscures the learner's contribution. Requiring documentation of the problem-solving process, such as interaction logs, reasoning summaries, or validation steps, exposes understanding. In this study, transcripts revealed whether students interpreted, tested, and refined results, providing richer evidence of learning than correctness alone.

***Recommendation***: Assess both product and process. Include a component that requires students to explain how they arrived at their solution and why the solution is trustworthy.

### 7.2 Treat Prompting as a Learnable Academic Skill

Student results improved as question structure and constraint specification became clearer. This indicates that interacting with AI effectively depends on analytical thinking rather than technical familiarity.

***Recommendation***: Explicitly teach prompt formulation. Provide examples of vague versus precise prompts and encourage students to refine questions iteratively. This transforms AI use from shortcut to skill.

### 7.3 Evaluate Verification, Not Just Answers

The strongest evidence of understanding appeared when students tested AI-generated outputs or identified mistakes. Verification behaviors demonstrated conceptual grasp even when initial answers were incorrect.

***Recommendation***: Design tasks where students must critique or validate solutions. For example, ask students to confirm correctness, identify potential errors, or justify why an output is reliable.

### 7.4 Design AI-Resilient Questions

Questions focused solely on explanation or definition can be answered directly by AI. In contrast, tasks requiring interpretation, comparison, or justification require student judgment.

***Recommendation***: Prioritize questions that require reasoning about results rather than reproducing information. For instance, ask students to evaluate alternatives, explain trade-offs, or apply concepts to novel contexts.

### 7.5 Use Transparency Instead of Policing

When AI use was allowed and documented, students demonstrated self-regulation and reflected on AI reliance. This suggests transparency reduces misconduct incentives.

***Recommendation***: Establish clear expectations for acceptable AI use and require acknowledgment rather than prohibition. Structure assessments so understanding, not concealment, becomes the primary concern.

### 7.6 Align Assessment with Professional Practice

In real-world settings, professionals use tools but remain accountable for outcomes. Assessments should mirror this environment by evaluating decision-making rather than isolated recall.

***Recommendation***: Frame AI as a tool students must supervise. Grade justification, interpretation, and critical evaluation as core competencies.

## 8. Conclusion and Future Work

This study examined how engineering students used ChatGPT during open-book examinations when its use was explicitly permitted and documented. Transcript analysis shows that generative AI reshapes assessment rather than simply simplifying it. Students progressed from retrieving answers to refining prompts and ultimately verifying outputs, with understanding most visible through questioning, testing, and justification of AI-generated results.

These findings indicate that answer correctness alone is no longer sufficient evidence of learning in AI-mediated environments. Assessment can instead reveal competencies such as judgment, verification, and responsible tool use when it focuses on reasoning about solutions rather than independent production. Transparent integration of AI shifted attention from rule enforcement toward self-regulation and reflective engagement. These behaviors parallel professional engineering practice, where computational tools are used but remain subject to critical oversight. Because the study is based on qualitative interpretation of transcripts from computing-related engineering courses, the findings should be interpreted as context-specific rather than broadly generalizable across all assessment settings.

Future work will examine how AI-mediated reasoning behaviors relate to longer-term learning outcomes and explore assessment designs that require students to critique or improve AI-generated solutions directly.

### Disclosure Statement

ChatGPT (OpenAI GPT-5.2) was used in a limited capacity to assist with language refinement and improvement of structural clarity. No AI tools were used to generate research findings, conduct data analysis, design the study, or create references. All conceptual development, data interpretation, and source verification were performed exclusively by the author.